\documentclass[10pt, conference, compsocconf]{IEEEtran}
\ifCLASSINFOpdf
\else
\fi
\usepackage{microtype}
\usepackage{graphicx}
\usepackage{booktabs} 

\usepackage{amsmath}
\usepackage{algorithmic}
\usepackage[linesnumbered]{algorithm2e}
\usepackage{color}
\usepackage[T1]{fontenc}

\usepackage[usenames,dvipsnames]{xcolor}
\usepackage{graphicx}

\usepackage{setspace}
\usepackage{comment}
\usepackage[most]{tcolorbox}

\usepackage{endnotes,microtype,xspace,graphicx,fancyvrb,multirow}

\usepackage{paralist}

\usepackage{mathptmx} 

\newcommand{\ignore}[1]{}
\usepackage{fancyhdr}
\usepackage[normalem]{ulem}
\usepackage[hyphens]{url}
\usepackage{longtable}
\usepackage{microtype}
\usepackage{graphicx}
\usepackage{caption}
\usepackage{subcaption}
\usepackage{color}

\usepackage[bookmarks=true,breaklinks=true,letterpaper=true,colorlinks,linkcolor=blue,citecolor=blue,urlcolor=red]{hyperref}
\usepackage[font=small]{caption}
\newcommand{\PCignore}[1]{}

\newcommand{\insertFigureNew}[3]{
    \begin{figure}[t]
\setlength{\abovecaptionskip}{-1pt}
\setlength{\belowcaptionskip}{-1pt}
        \centering
        \includegraphics[width=#3\linewidth]{figs/#1.pdf}
        \caption{\small #2}
        \label{fig:#1}
    \end{figure}
}
\newcommand{\insertFigureNewPNG}[3]{
    \begin{figure}[t]
\setlength{\abovecaptionskip}{-1pt}
\setlength{\belowcaptionskip}{-1pt}
        \centering
        \includegraphics[width=#3\linewidth]{figs/#1.png}
        \caption{\small #2}
        \label{fig:#1}
    \end{figure}
}

\newcommand{\insertNewWideFigurePNG}[3]{
    \begin{figure*}[h]
    \setlength{\abovecaptionskip}{-1pt}
    \setlength{\belowcaptionskip}{-4pt}
        \centering
        \includegraphics[width=#3\linewidth]{figs/#1.png}
	    \vspace{-2mm}
	    \bigskip
        \caption{\small #2}
    	\vspace{-2mm}
        \label{fig:#1}
    \end{figure*}
}

\newcommand{\squishlist}{
 \begin{list}{$\bullet$}
  { \setlength{\itemsep}{0pt}
     \setlength{\parsep}{3pt}
     \setlength{\topsep}{3pt}
     \setlength{\partopsep}{0pt}
     \setlength{\leftmargin}{1.5em}
     \setlength{\labelwidth}{1em}
     \setlength{\labelsep}{0.5em} } }

\newcommand{\squishlisttwo}{
 \begin{list}{$\bullet$}
  { \setlength{\itemsep}{0pt}
     \setlength{\parsep}{0pt}
    \setlength{\topsep}{0pt}
    \setlength{\partopsep}{0pt}
    \setlength{\leftmargin}{2em}
    \setlength{\labelwidth}{1.5em}
    \setlength{\labelsep}{0.5em} } }

\newcommand{\squishend}{
  \end{list}  }

\newcommand{\SSR}[1]{\textcolor{teal}{Srinivas: #1}}

\begin{document}
%
\title{Impact of RoCE Congestion Control Policies on Distributed Training of DNNs}



\author{\IEEEauthorblockN{Tarannum Khan\IEEEauthorrefmark{1}, Saeed Rashidi\IEEEauthorrefmark{2}, Srinivas Sridharan\IEEEauthorrefmark{3},
Pallavi Shurpali\IEEEauthorrefmark{3},
Aditya Akella\IEEEauthorrefmark{1}, and
Tushar Krishna\IEEEauthorrefmark{2} 
}
\IEEEauthorblockA{\IEEEauthorrefmark{1}The University of Texas at Austin, 
Austin, USA\\
}
\IEEEauthorblockA{\IEEEauthorrefmark{2}Georgia Institute of Technology, 
Atlanta, USA\\
}
\IEEEauthorblockA{\IEEEauthorrefmark{3}Meta, 
Menlo Park, USA\\
\textit{ tarannum.khan@utexas.edu, saeed.rashidi@gatech.edu, 
ssrinivas@fb.com, tushar@ece.gatech.edu}}}


%


\maketitle

\begin{abstract}
RDMA over Converged Ethernet (RoCE) has gained significant attraction for datacenter networks due to its compatibility with conventional Ethernet-based fabric. However, the RDMA protocol is efficient only on (nearly) lossless networks, emphasizing the vital role of congestion control on RoCE networks. Unfortunately, the native RoCE congestion control scheme, based on Priority Flow Control (PFC), suffers from many drawbacks such as unfairness, head-of-line-blocking, and deadlock. Therefore, in recent years many schemes have been proposed to provide additional congestion control for RoCE networks to minimize PFC drawbacks. However, these schemes are proposed for general datacenter environments.
In contrast to the general datacenters that are built using commodity hardware and run general-purpose workloads, high-performance distributed training platforms deploy high-end accelerators and network components and exclusively run training workloads using collectives (All-Reduce, All-To-All) communication libraries for communication. Furthermore, these platforms usually have a private network, separating their communication traffic from the rest of the datacenter traffic. Scalable topology-aware collective algorithms are inherently designed to avoid incast patterns and balance traffic optimally. These distinct features necessitate revisiting previously proposed congestion control schemes for general-purpose datacenter environments. In this paper, we thoroughly analyze some of the state-of-the-art RoCE congestion control schemes (DCQCN, DCTCP, TIMELY, and  HPCC) vs. PFC when running on distributed training platforms. Our results indicate that previously proposed RoCE congestion control schemes have little impact on the end-to-end performance of training workloads, motivating the necessity of designing an optimized, yet low-overhead, congestion control scheme based on the characteristics of distributed training platforms and workloads.




\end{abstract}


%
\IEEEpeerreviewmaketitle

\section{Introduction}\label{sect:intro}
Deep Neural Network (DNN) models are gaining significant attention due to their wide applicability across different domains such as vision \cite{ResNet,Behnam1,Behnam2}, language modeling \cite{Transformer1T}, sensor networks \cite{AfshinDistributedInference}, and recommendation systems \cite{DLRM}. DNN workloads first need to be trained on existing samples to achieve high accuracy.  
However, as the number of training samples and DNN model sizes increase, training is becoming increasingly challenging, requiring several days/weeks to be trained \cite{NVidiaSwitch,CommBottleneck2,blueconnect}. This is especially true for DNNs such as recommendation models~\cite{DLRM} that need to be constantly trained using new data samples produced daily.

Distributed training is used in practice today as a solution to reduce the training time by distributing the training task across multiple accelerators, aka \emph{Neural Processing Units} (NPU - a term that abstracts any specific accelerator type (e.g., TPU, GPU, FPGA)). However, distributed training comes at the expense of communication overhead between NPUs to synchronize model gradients and/or activation depending on the \emph{parallelization strategy} of the distributed training. This communication is usually handled through a series of  synchronous and multi-peer patterns, called \emph{collective communication patterns} (e.g., All-Reduce) \cite{NVidiaSwitch,Themis}. Scalable topology-aware collective algorithms are designed to avoid in-cast patterns and balance traffic optimally. 

To achieve maximum performance, in recent years specialized distributed training platforms have been built, both by cloud providers~\cite{googleCloudTpu,FBDLRMPlatform,microsoft_brainwave} and accelerator vendors~\cite{NvidiaH100,HabanaPtP}. 
These platforms have distinct characteristics that set them apart from the conventional data-center environments:

First, compared to the traditional data-center setting that relies on commodity hardware, distributed training platforms are built using high-end compute and network components \cite{FBDLRMPlatform,dgx2,TACCL}. In such an environment, it is typical for each NPU to have a dedicated high-BW Network Interface Card (NIC), supported by direct remote memory access (RDMA) to distant NPUs. This provides enormous network BW per server node, given that each server node hosts multiple NPUs (up to 16 per server node \cite{dgx2}).

Second, to mitigate the communication overhead, distributed training platforms employ dedicated networks that separate training traffic from the rest of the datacenter traffic \cite{FBDLRMPlatform}. In this case, each server node has a dedicated NIC per CPU socket to connect to the datacenter network, while NPUs across different server nodes form their private network using their dedicated NICs \cite{FBDLRMPlatform}.

Third, due to the growing size of DNN models and the significantly growing size of training datasets that are generated daily, training platforms are often scheduled to perform only one training job at a time for the critical DNN workloads (e.g., recommendation models) to minimize the training time and enable training over larger datasets \cite{hotiPaper}. 

As a result of these unique characteristics, it is necessary to revisit the networking stack and identify whether current state-of-the-art networking protocols are optimal for such platforms. In particular, we focus on \emph{RDMA over Converged Ethernet} (RoCE) protocol \cite{RoCE} due to its compatibility with current Ethernet-based fabric and widespread usage on distributed training platforms \cite{FBDLRMPlatform}. Recent works have shown the importance of congestion control on RoCE to achieve maximum performance \cite{DCQCN,Rogue,Timely,HPCC}. This is because the RDMA protocol is more efficient on lossless networks, which is not natively supported on Ethernet-based fabrics \cite{RoCE}. 
To address this issue and achieve near-lossless network guarantees, baseline RoCE enforces congestion control at the link layer through the Priority Flow Control (PFC) mechanism. In this case, once the receiver-side buffer occupancy crosses a threshold, the receiving NIC generates a PAUSE frame and sends it to the sender who  stops sending packets until further notified by the receiver.

However, as previous works have shown, the PFC mechanism suffers from many drawbacks in conventional data-center environments, including unfairness, head-of-line-blocking, and deadlock \cite{DCQCN}. Therefore, various works have proposed additional NIC- and network-based congestion control techniques on top of PFC enforced to achieve high-performance and lossless features while minimizing the PAUSE frames generated by PFC. \cite{DCQCN,Rogue,Timely,DCQCN,HPCC}.

In this paper, we study the effect of these state-of-the-art congestion control mechanisms -- proposed for general data-center workload platforms/environments -- and compare them with baseline PFC in the context of distributed training workloads and platforms. We study the impact on end-to-end training time and explain the observations. 

We make the following contributions:
\squishlist
    \item To the best of our knowledge, this is the first work that evaluates the effect of different congestion control schemes on distributed training.
    \item We developed a simulator using ASTRA-Sim \cite{astrasim} and NS3 \cite{AlibabaNS3}. We model state-of-the-art training workloads and platforms in ASTRA-Sim \cite{astrasim} and integrate with NS3 \cite{AlibabaNS3} to model the RoCE network and its different congestion control schemes.
    \item We provide a detailed analysis of the effect of each state-of-the-art congestion control scheme (i.e., Baseline PFC, DCQCN, DCTCP, Timely, and HPCC) for both single collective communication micro-benchmarks and end-to-end training time of the DLRM workload~\cite{DLRM}.
    \item We show that different state-of-the-art RoCE congestion control schemes have little impact on the end-to-end training performance. 
    \item Based on our analysis, we provide directions for designing an optimized yet low-overhead congestion control scheme tuned for distributed training. 
\squishend

\section{Background}\label{sec:background}
\subsection{Collective Communication Patterns}\label{subsec:collPatterns}
Collective communications are the most frequent communication pattern observed in distributed training \cite{NVidiaSwitch,blueconnect,TopologyPaper}. Collective communication patterns refer to a set of collaborative and synchronous data exchange between NPUs for a specific purpose~\cite{hotiPaper,NVidiaSwitch}. The most frequently used pattern in distributed training is \emph{All-Reduce}. In All-Reduce, each NPU initially has a data buffer, and the end goal is to \emph{reduce} the data buffer of all NPUs and replicate the results on all NPUs~\cite{collective1,BLink}. \emph{All-Reduce} can be broken into executing two consecutive communication steps: i) \emph{Reduce-Scatter} followed by ii) \emph{All-Gather} \cite{blueconnect}. In the Reduce-Scatter phase, each NPU obtains a portion of the globally reduced data buffer, while in All-Gather, each NPU broadcasts their data buffer to all other NPUs~\cite{hotiPaper}.

In addition to All-Reduce, All-To-All is another collective pattern observed for some training workloads such as DLRM recommendation models \cite{DLRM}. In All-To-All, each NPU sends a specific portion of each data buffer to each NPU. 


\subsection{Collective Communication Algorithms}
\textbf{All-Reduce.} To handle the collective patterns discussed in \autoref{subsec:collPatterns}, many different algorithms have been proposed in recent years. For example, basic All-Reduce algorithms include: ring-based \cite{collective1}, tree-based \cite{BLink}, halving-doubling \cite{EFLOPS}, etc. Each basic algorithm is optimal for certain physical topologies. For example, the ring-based algorithm is optimized for physical ring topology, while direct All-Reduce works well for physical switch-based topologies. \cite{hotiPaper}.

In the basic direct All-Reduce algorithm among P NPUs, the data buffer is split into P segments, and each NPU sends its i'th segment (0 $\leq$ i < P) to the NPU id \#i at the same time. Therefore NPU id \#i receives all i'th segments of all NPUs, which then reduce with its local i'th segment (Reduce-Scatter). In the next phase, each NPU broadcasts its locally reduced segment to all other NPUs at the same time (All-Gather) \cite{hotiPaper}.

For heterogeneous network topologies with multiple levels of the network, multi-stage hierarchical All-Reduce algorithms have been proposed~\cite{blueconnect,BLink}. In this case, All-Reduce is broken into a sequence of Reduce-scatter operations starting from the first level and going all the way to the last network level. This is then followed by issuing All-Gathers in the reverse order of Reduce-Scatter. The Reduce-Scatter/All-gather algorithm for each stage is a basic collective algorithm discussed earlier (e.g., ring-based, halving-doubling, etc.). Doing so reduces network traffic as data goes to the next network level, which is desired since later network levels have less bandwidth than the first levels. 

\textbf{All-To-All.} Like All-Reduce, the All-To-All pattern can also be implemented in multiple ways. However, the most common algorithm is direct All-To-All, where each NPU sends its messages to other destination NPUs at the same time \cite{hotiPaper,nccl,oneccl}. The reason is that, unlike All-Reduce, hierarchical All-To-All does not reduce the network traffic as data goes to the next network level.

\subsection{DLRM Workload}
Recommendation models are one of the most important class of DNN workloads that need to be trained constantly with a huge amount of user data generated daily. Therefore, they serve as a suitable target for our studies since they require maximum training performance to keep up with the growing user training data \cite{FBDLRMPlatform}. 

DLRM \cite{DLRM} is the most common DNN model used for recommendation systems. DLRM combines user-obtained sparse features (e.g., pages the user previously liked) and dense features (e.g., age) and produces the \emph{Clickthrough rate (CTR)} for a user to view a specific Ad. It consists of three main parts: bottom MLP, embedding layer, and top MLP. Bottom MLP is used to process dense features while embedding tables represent sparse features. Sparse features are looked up in the tables, and then the output of embedding tables interacts with each other (e.g., using dot product) to generate the output of the embedding layer. The embedding layer and bottom MLP layers can run in parallel, and their outputs are then concatenated to feed the top MLP layers to predict CTR.

The two most common parallelization strategies in DNN training is \emph{data-parallel} and \emph{model-parallel}. In data-parallel, the model is replicated across NPUs, and each NPU works on separate mini-batches. This strategy requires an All-Reduce between model gradients computed by all NPUs before updating the model and starting the new training iteration. In model-parallel, the model is split across NPUs, but all NPUs work on the same minibatch. Therefore it requires collective communications between NPUs to exchange output activations/input gradients of each model-parallel layer during forward-pass/backpropagation. The exact model-parallel communication type depends on the layer type.

The standard parallelization strategy of DLRM is to distribute the MLP layers in a data-parallel manner, while the embedding tables are split across all NPUs due to their huge sizes, forming a model-parallel split \cite{DLRM,FBDLRMPlatform}. Therefore, DLRM requires All-Reduce collectives for data-parallel MLP layers, and All-To-All for forward-pass/backpropagation of embedding layer \cite{DLRM}.

\subsection{RoCE Congestion Control Schemes}
We evaluated our target workloads on the following congestion control policies:
\subsubsection{PFC only}
In this case, we did not use any congestion control algorithm and used only pause frames to control the congestion \cite{RoCE}. When the bytes in the queues are more than a threshold value, pause frames (PFCs) are triggered, and when these PFCs reach the sender side, the sender stops sending more packets until the sender receives resume frames sent by the switch.
\subsubsection{DCQCN}
DCQCN \cite{DCQCN} is inspired by QCN \cite{QCN} and DCTCP \cite{DCTCP}. Once the queue length surpasses the ECN threshold, a congestion notification packet is generated from the receiver of the ECN-marked packet to inform the sender to reduce the flow rate. If the sender receives no feedback for a fixed time unit, the sender starts to increase the rate. Initially, it increases the rate rapidly in the fast recovery phase, and then it additively increases the rate. DCQCN starts at line rate. DCQCN has many parameters that need to be tuned for better performance.
\subsubsection{DCTCP}
DCTCP \cite{DCTCP} achieves low latency and high throughput requirements by reacting to congestion in proportion to congestion. DCTCP uses a simple marking scheme at switches. It marks packets by setting the ECN flag at the switches if the queue occupancy exceeds the threshold. The sending window is reduced at the sender side depending on the reduction factor, alpha. DCTCP was originally proposed for the TCP network. We used the same algorithm developed for datacenter TCP traffic but used it for RoCE v2 as is done in HPCC \cite{HPCC} with the starting rate as line rate so that it can be equivalently compared with other congestion algorithms which are starting at line rate.
\subsubsection{TIMELY}
TIMELY \cite{Timely} is a congestion control mechanism that measures round trip time (RTT) delay and accordingly decides to reduce or increase the rate at the sender side. If the RTT is less than Tlow, the rate is increased additively. If the RTT is greater than Thigh, the rate decreases multiplicatively, and if RTT is between Tlow and Thigh, the rate increases or decreases based on the gradient.
\subsubsection{High Precision Congestion Control (HPCC)}
Earlier algorithms were doing congestion control based on ECN or delay. But, HPCC does congestion control through the in-network telemetry, INT feature of its switching ASIC where it inserts information such as timestamp, queue length, etc. When the receiver receives its packet, it copies this metadata to the ack and sends the ack packet to the sender to take action depending on the metadata information in the ack packets. HPCC also controls inflight bytes by using a window and varying it according to the INT header information. HPCC does some more interesting optimization to monitor congestion closely \cite{HPCC}.
\subsubsection{HPCC-PINT}
In HPCC, there is an INT overhead for every packet as each switch adds this information to the packet. A data center topology with five hops with the only addition of two values per switch requires 48 bytes of overhead or 4.8 percent of a 1000 bytes packet. The HPCC problem of overhead per packet is solved using PINT with HPCC, which uses only 8 bits. HPCC-PINT \cite{HPCCPINT} achieves this by not giving feedback per packet but depending on a parameter. Hence feedback can sometimes be delayed, especially for shorter flows.

\section{Methodology}\label{sec:methodology}
\subsection{Simulation Methodology}


We use the ASTRA-SIM simulator \cite{AstraSimGithub}\cite{astrasim} to model the communication scheduling of deep learning (DL) training workloads (both individual collectives and actual DL models).
ASTRA-SIM is an open-source cycle-level simulator for DL training platforms developed by Georgia Tech, Meta, and Intel~\cite{astrasim}, validated against current training platforms. \autoref{fig:astrasim} shows an overview of ASTRA-SIM. It comprises three layers. The \textit{workload layer} takes the DNN model descriptions as an input, in terms of compute times and communication operations per layer, and simulates the training loop. 
In this work, we use NVIDIA V100 GPU profiling to obtain compute times.
For simulating the communication operations, the workload layer calls the \textit{system layer} communication APIs that implement diverse collective communication patterns via various algorithms (e.g., ring-based, halving-doubling, and so on). To accurately model the collective communications, the system layer breaks the algorithm into a series of send/recv operations that should be modeled via the \textit{network layer}. ASTRA-sim provides a network API~\cite{hotiPaper} to plug in diverse network backends. We integrated the Alibaba NS3 simulator~\cite{AlibabaNS3} as the network layer in this work. NS3 models different state-of-the-art RoCE congestion control (CC) schemes including: Baseline PFC \cite{RoCE}, DCQCN \cite{DCQCN}, Timely \cite{Timely}, DCTCP \cite{DCTCP} and HPCC \cite{HPCC}. We use these CC algorithms for our comparative analysis. To the best of our knowledge, this is the first integrated simulator that models distributed training workloads on RoCE networks.

\insertFigureNew{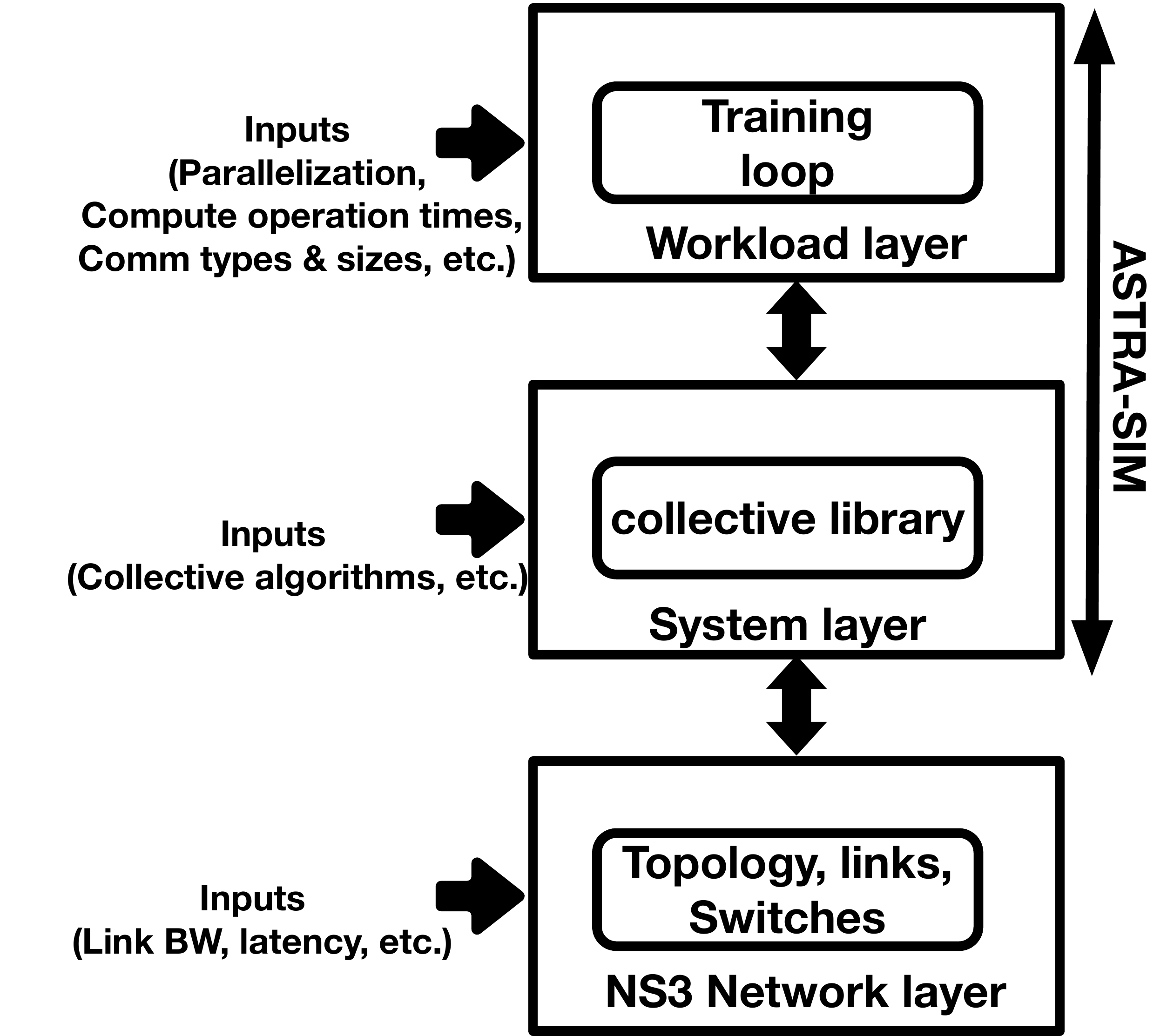}{The Simulation methodology architecture}{0.69}

\subsection{Target Platforms}
We model the distributed training platform similar to Mudigere et al.~\cite{FBDLRMPlatform}.
\autoref{fig:fbtopo} presents the topology architecture. In this case, each server nodes host 8 GPUs, locally connected to six NVSwitches (scale-up network) within a server node. Each GPU has a dedicated NIC connecting it to the top-of-the-rack (TOR) switch (scale-out network). There are two such server nodes within a rack. The inter-rack connectivity is enabled by connecting TOR switches to spine switches as shown in \autoref{fig:fbtopo}.
The connectivity is a 1:1 full subscription non-blocking CLOS topology. This paper simulates up to sixteen racks with eight spine switches.

\autoref{table:designParams} summarizes the topology design parameters. Note that in this design, the GPU network is separated from the general datacenter traffic for maximum training performance\footnote{In this case, the connectivity of each server node to the rest of the datacenter is enabled via the CPU NIC (not modeled in this paper).}\cite{FBDLRMPlatform}. 

\insertFigureNewPNG{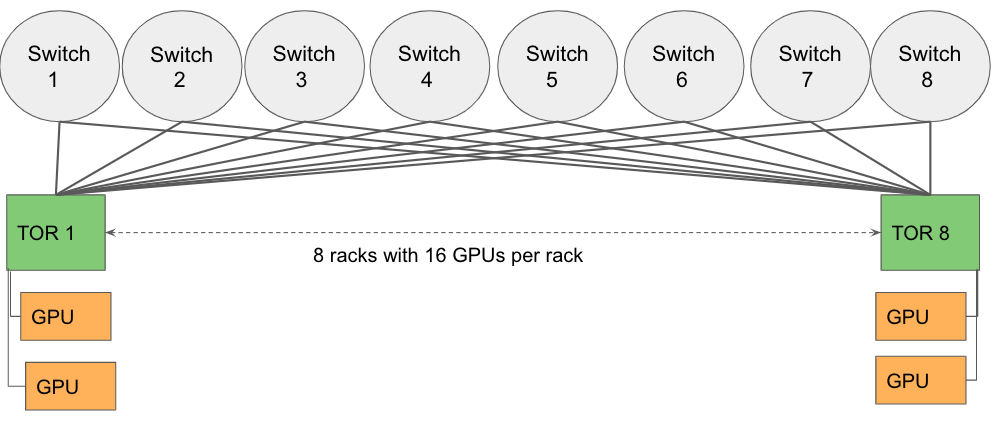}{The two-level Clos topology commonly used in datacenters.}{1}

\begin{table}[]
\caption{Topology design parameters}
\label{table:designParams}
\resizebox{0.8\columnwidth}{!}{%
\begin{tabular}{|l|l|}
\hline
\textbf{Parameter}                      & \textbf{Value}             \\ \hline
Scale-up (NVlink) link BW      & 200 GBps (total) \\ \hline
Scale-up (NVlink) link latency & 25 ns             \\ \hline
NIC-to-TOR link BW             & 200 Gbps (total)  \\ \hline
NIC-to-TOR link latency        & 500 ns            \\ \hline
TOR-to-Spine link BW           & 200 Gbps (total)  \\ \hline
TOR-to-Spine link latency      & 500 ns            \\ \hline
TOR Switch buffer size         & 32 MB             \\ \hline
Spine switch buffer size       & 32 MB              \\ \hline
\end{tabular}%
}
\end{table}

\subsection{Target Workloads}\label{subsec:targetWorkload}
We use incast, single collective micro-benchmark workloads (i.e., All-Reduce and All-To-All), and real end-to-end training workloads to compare different CC algorithms. The goal is to first understand CC behaviours for simpler communication patterns and then extend it to the full end-to-end training performance. For real training workloads, we pick distributed training of the DLRM \cite{DLRM} recommendation model since it is one of the critical workloads that need constant training over newly generated samples and hence, requires maximum training performance \cite{FBDLRMPlatform}. The DLRM model size is similar to \cite{hotiPaper}. \autoref{table:modelParams} shows our DLRM description.

\textbf{Multi Tenancy. }As discussed in \autoref{sect:intro}, to model the critical training workloads, we assume only a single training workload is under execution on our target platforms. This is inline with many real system scenarios where maximum performance is required to keep pace with the constant need to train the network with newly generated training data produced daily \cite{FBDLRMPlatform,hotiPaper,NVidiaSwitch,saeedACE}.

\begin{table}[]
\centering
\caption{DLRM Model parameters}
\label{table:modelParams}
\resizebox{0.6\linewidth}{!}{%
\begin{tabular}{|l|l|l|}
\hline
\textbf{Model Parameters}            & \textbf{Value}   \\ \hline
Size of embedding data-type & 16 bits \\ \hline
Pooling factor              & 60 \\ \hline
Top MLP layers              & 10+2 \\ \hline
Bottom MLP layers           & 5+2 \\ \hline
Dense features              & 1600 \\ \hline
Top MLP layer size          & 2048 \\ \hline
Bottom MLP layer size       & 1024 \\ \hline
Sparse features             & 64 \\ \hline
Embedding dimension         & 64 \\ \hline
\end{tabular}%
}
\end{table}

\subsection{Target Collective Communication Algorithms}\label{subsec:targetCollective}
For All-Reduce, we use two different versions of All-Reduce: i) \emph{1D}: which is a basic direct All-Reduce algorithm, and ii) \emph{2D}: which is a BW-efficient hierarchical algorithm as described in \autoref{sec:background}. This means that on our target topology, an All-Reduce is broken into: Reduce-Scatter within each server node through scale-up network (NVLink), followed by Reduce-Scatter between GPUs of the same id across different server nodes through scale-out NICs, followed by All-Gather in the reverse order (four steps in total). Since our topology is switch-based at all levels, the basic All-Reduce algorithm on each stage is a direct algorithm as it is shown to be efficient on switch-based networks \cite{hotiPaper}.

For All-To-All, we perform the direct All-To-All algorithm described by Rashidi et al. \cite{hotiPaper} and \autoref{subsec:collPatterns}. 

Also, to utilize all network levels simultaneously, we break each collective into four equally sized chunks and then process them in a pipeline manner, similar to \cite{GC3}.


\subsection{Metric of Evaluation}
We use various metrics to evaluate the performance and efficiency of CC algorithms for microbenchmark and real workload cases. We mainly focus on network stats such as end-to-end completion time, switch buffer occupancy, and \#of PFC PAUSE frames for microbenchmark communication workloads.

For real workloads, we focus on mainly workload-related stats such as end-to-end training iteration time that can be decomposed into \emph{total compute times} + \emph{exposed communication time}. Exposed communication time is when the training loop is forced to stop because it needs to wait for the communication to be finished. In other words, it shows the amount of communication time that cannot be overlapped with compute.
\section{Results \& Analysis}
\insertNewWideFigurePNG{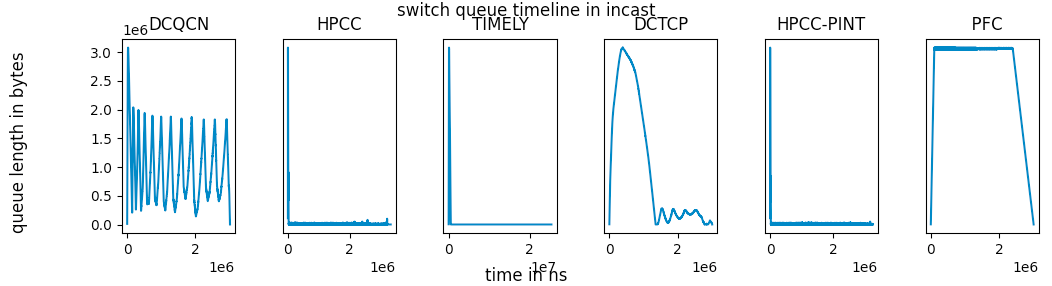}{Queue length timeline in incast}{0.9}
\subsection{Single-switch Incast micro-benchmark}
For studying incast, we used a simple topology where 8 GPUs are connected to a switch. The switch has a buffer size of 32 MB. All the GPUs are connected to a switch with a link of bandwidth 200 Gbps and 500 ns latency. Seven of the GPUs are sending 10MB data to GPU 0. In all the figures in result, queue length in bytes is the sum of bytes in all the queues in a switch at a time. We study how the egress buffer queue is varying in all the algorithms. Note that in this case, the buffer size will be saturated first when compared with the link connecting GPUs and the switch as queue buffer is a scarce resource compared to the link bandwidth. Hence utilization of the queue such that we are not underutilizing the bandwidth becomes an important criteria for different congestion control algorithm to perform well. We observe similar performance for all CC algorithms. 

\subsubsection{PFC only}
For the case where only PFC \cite{RoCE} is enabled without any additional congestion control, we see there is a queue build up throughout the flow time as shown in \autoref{fig:incast}. This is because, once the queue is build and reaches the threshold value of a switch queue, PFCs are triggered and whenever the queue length is dropped below the threshold, resume is triggered which allow again more packets to pass through the link and into the switch. Hence, maximum queue length is utilized and also the link bandwidth is utilized efficiently. In case of using only PFC, there are no complex computation which needs to be done like in other congestion control policies. Hence, this also saves some time. But this advantage, comes with a cost of a lot of PFC’s to get triggered. There are a lot of problems with PFC like head-of-line blocking, deadlocks and unfairness \cite{DCQCN}. Because of these reasons, various congestion control were introduced to avoid PFC production as much as possible.

\subsubsection{DCQCN}
The first congestion control algorithm, we will study is Datacenter QCN, DCQCN \cite{DCQCN} which can be seen in \autoref{fig:incast}. DCQCN is a fluid model with various parameters in switch to be set. We tuned the parameter in every experiment to test against the best case of DCQCN. In this case, there are no PFC generated and we get better link utilization with less queue buffer usage than the previous case where a lot of PFCs were generated. Hence, we have low latencies and also eliminating the problem which comes with PFC. In the incast case here, first time there is a shoot up in the switch queue, as there is a delay in the initial rate reduction as the sender will have to wait for the first congestion notification packet to be  received from the switch. After, that we can see the queue buffer has stabilised to be within a range. 

\subsubsection{DCTCP}
The next algorithm we studied is DCTCP \cite{DCTCP}. 
As we can see in  \autoref{fig:incast}, the queue is being built up initially but then due to the ecn notification received from the switches the sender decreased there window size and the eventually the queue size becomes very small. Here, also the PFC trigger point is not reached and no PFCs are being produced.

\subsubsection{TIMELY}
As we can see in \autoref{fig:incast}, TIMELY reduces the rate heavily initially. As TIMELY receives delayed RTT initially because of the switch queue buildup, it starts aggressively reducing the rate of the senders. As the rate of all the seven GPUs sending to the zeroth GPUs get reduced together, the switch buffer usage becomes very low. With the low queue buffer usage, the bandwidth is also underutilized and the latency is worst for TIMELY. We tried various parameters but we were getting almost the same latency, so we used the values provide in the TIMELY paper. Hence, for large flow, very less queue buildup may not be a good option as there are no shorter flows here which will get affected by the queue build ups. As we can see in DCQCN, which has the least latency, a queue build up which is also getting consumed quickly is an ideal buffer pattern for large flows, so that the bandwidth can also be utilized in this scenario and PFCs are not produced. In TIMELY too for this case, PFCs are not generated. 
\subsubsection{HPCC}
As we can see in  \autoref{fig:incast}, HPCC through the INT Header is continuously monitoring the queue length when it start receiving it acks and hence, continuously starts decreasing the window size until the queue length of the switch is minimal, which is the aim of HPCC. In this case, also no PFCs are produced. It is worth noting every packet there is an INT overhead as each switch adds this information into the packet. Hence, we are actually transferring more data than in other congestion control schemes and this may increase flow completion time. It also increases processing latency at switches \cite{HPCCPINT}. 

\subsubsection{HPCC-PINT}
The solution to the HPCC problem of overhead per packet is solved using PINT along with HPCC which uses only 8 bits and have similar results and time series of queue length when compared with HPCC as seen in \autoref{fig:incast}. HPCC-PINT achieves this by not giving feedback per packet but depending on a parameter, resulting in delayed feedback. This delayed feedback will worsen results for shorter flows but since in our study we are focusing on collectives which are generally longer flows, the results are better as the overhead per packet is reduced.
\insertFigureNewPNG{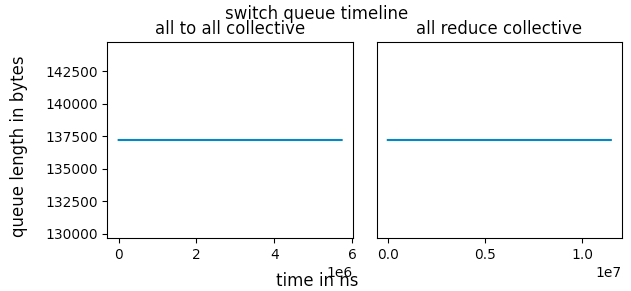}{Queue timeline for All-To-All and All-Reduce collective}{0.8}

\insertFigureNewPNG{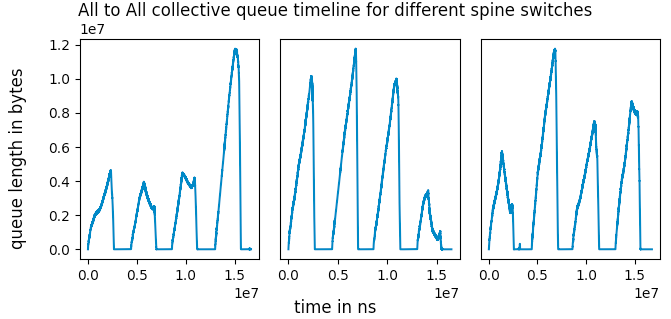}{Effect of ECMP as observed from queue length timeline for different Spine switches for the same All-To-all Collective. }{0.8}

\insertNewWideFigurePNG{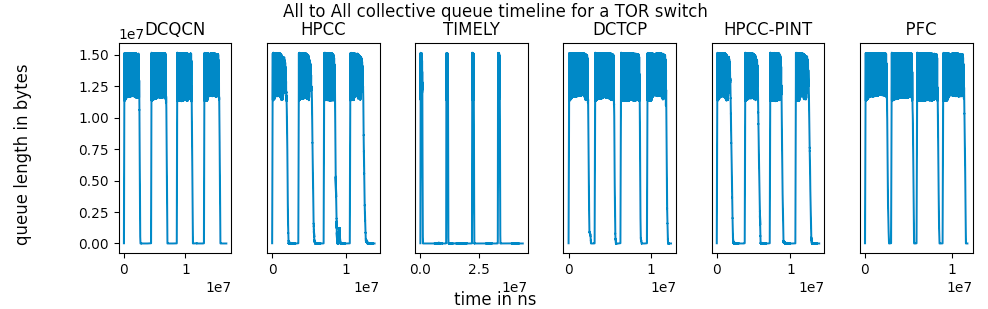}{Queue timeline for a TOR switch for different CC for All-To-All collective}{0.8}
\insertNewWideFigurePNG{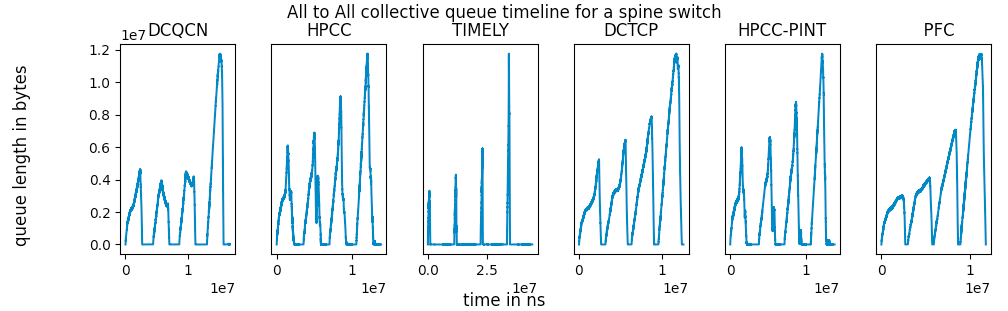}{Queue timeline for a Spine switch for different CC for All-To-All collective}{0.8}

\subsection{Single-switch Collectives Micro-benchmark}
We studied All-Reduce and All-To-All workload for one switch connected to 8 GPUs and scaled it to one switch connected to 128 GPUs. We use the bandwidth of 200 Gbps and 500 ns latency and a switch of buffer size 32 MB. We also studied the impact of different congestion control algorithms for these workloads. Due to the nature of how these collectives are designed, we observed there was no congestion in this network and all the congestion control algorithms have similar completion times.

Let’s consider All-To-All where each node is sending data to all other nodes. If the total collective data to be communicated is 10MB, then in the case where there are 8 GPUs, each GPU will send 1.25MB data to each of the other GPU. In the case of incast all of the 7 GPUs were sending data to the same queue of the switch. Thus, one switch queue was receiving 7*10MB of data. In All-To-All, as the data is distributed among different GPUs, it is also distributed among different switch queues. Hence, each of the switch queue will received 8 * 1.25MB of data from the 8 GPUs which is not enough to cause congestion, therefore all the algorithms produce similar latency. We have not added simulation results for less than 128 GPUs considering space limits as we see similar results as in \autoref{fig:constq}, when sending 10 MB of data from all the nodes in All-Reduce and All-To-All for 8 GPUs. We observe that the sum of all the switch queue lengths is constant with respect to time showing no queue buildup and no congestion for all the congestion control policies. 
When this is scaled from 8 GPUs to 128 GPUs and the collective communication is scaled up to 128MB, similar results are obtained due to the same pattern of data distribution. The same pattern is observed in All-Reduce as well. Hence, in \autoref{fig:constq}, we can see that there is no queue buildup and no congestion. As there were no congestion, we also observed there were no PFC generation for all the congestion control algorithms.

\subsection{Two-level CLOS topology}
In this section, we study the effect of congestion on the common topology used in the datacenter, which is a two-level CLOS topology. This topology is modeled similar to \cite{FBDLRMPlatform} and shown in \autoref{fig:fbtopo}.
We studied All-Reduce, All-To-All, and the DLRM workload for different congestion control policies in this topology setting.

\subsubsection{All-To-All Collective}
\insertNewWideFigurePNG{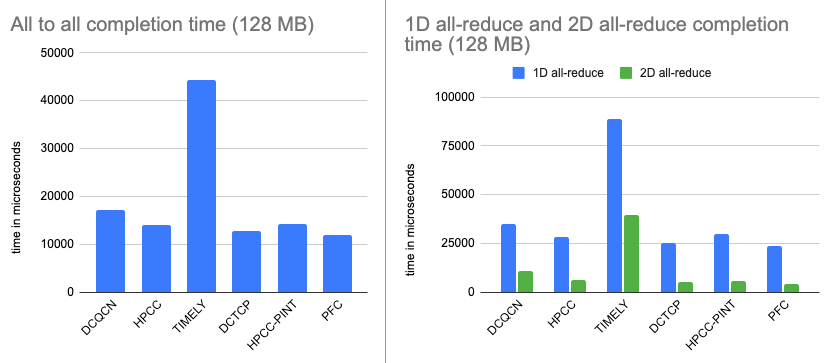}{Completion time for collectives}{0.6}
\insertNewWideFigurePNG{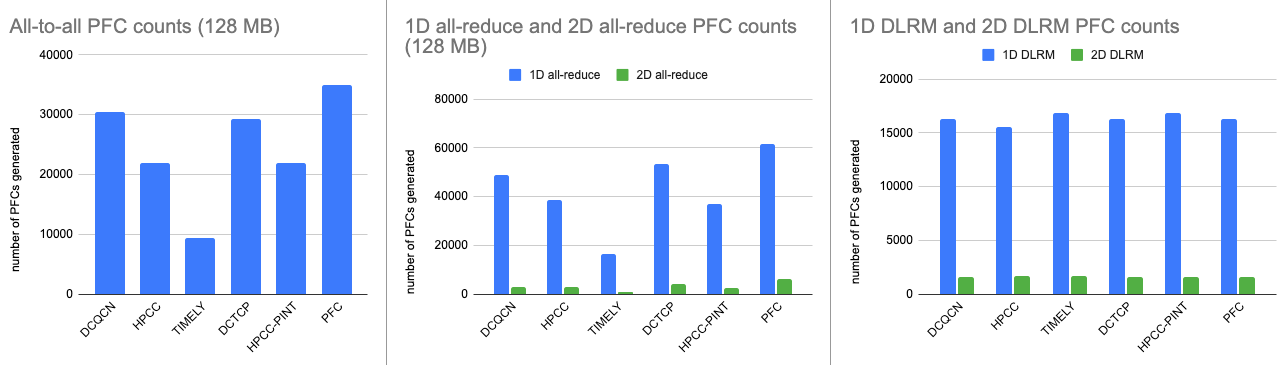}{PFC counts for workloads}{0.9}
\autoref{fig:a2a_TOR} and \autoref{fig:a2a_spine} depicts queue timeline for ToR and Spine switches for all the congestion control algorithms for All-To-All collective. We observe four peaks in \autoref{fig:a2a_TOR} and \autoref{fig:a2a_spine} as we have split the data into four chunks which means we are completing four All-To-All flows from all the GPUs, one by one. As each chunk issues, all sends in one burst and then waits for completion of all its messages, therefore, we see four queue length peaks in \autoref{fig:a2a_TOR} and \autoref{fig:a2a_spine}. \autoref{fig:ECMP} depicts the queue timeline for three different Spine switches for the same All-To-All collective. From  \autoref{fig:ECMP}, we can observe that the Spine switches have different queue build-ups simultaneously for the same All-To-All collective flow. As the same Spine switch can be responsible for sending data to different TORs depending on the ECMP hash, in some of the switches, there could be more queue build-up compared to other switches at a point in time, as seen in \autoref{fig:ECMP}. Hence, we can see queue build-up and congestion at different times in different Spine switches, which is depicted in \autoref{fig:ECMP} from the perspective of different switches and \autoref{fig:a2a_spine} from the perspective of different congestion control algorithms. For TOR switches in those communication periods, the queue lengths are in a constant range as seen in \autoref{fig:a2a_TOR} which indicates that there is not much congestion in the TOR switches. 

All the congestion control algorithms have a similar queue build-up and latency except for TIMELY, which can be seen in \autoref{fig:DLRM_CT} where we perform an All-To-All collective for 128 MB data. Similar results were observed for other data sizes as well. TIMELY being a rate-based algorithm, after receiving RTT delay way more than expected initially, it reduced the rate heavily and hence started underutilizing the bandwidth. We also suspect another reason for TIMELY poor performance could be non-optimal parameters. We used the parameters mentioned in the TIMELY paper and played with parameters close to the ones mentioned in the TIMELY paper. However, we still observed poor performance when using TIMELY compared to other algorithms.


Overall, while the PFC-only scheme generates the maximum number of PFCs, as shown in \autoref{fig:DLRM_pfc}, we observe it achieves the equal or best performance compared to other CC schemes (except TIMELY). Because we have long flows, PFC does not degrade the BW utilization. 

\subsubsection{All-Reduce Collective}
As in All-To-All, All-Reduce also has congestion because of ECMP, as explained in the previous section. For All-Reduce, we see a similar queue buildup for TOR and spine switches for all the congestion control algorithms as in All-To-All collective.
For All-Reduce completion time as seen in \autoref{fig:DLRM_CT}, we ran 128 MB All-Reduce for 128 GPUs. Similar results were observed for other data sizes as well. Similar to all-to-all, in all-reduce also because of the same reasons, the performance of TIMELY is the worst among all congestion control algorithms. All other congestion control algorithms have similar performance compared to only PFCs with no congestion control algorithm. 

Additionally, we compared how 1D and 2D collectives fare in this topology set. 1D All-Reduce is doing All-Reduce throughout 128 GPUs. 
2D All-Reduce means first doing All-Reduce in each server node and then doing All-Reduce among other GPUs in the servers, hence sending less data into the TOR and Spine switches in 2D All-Reduce compared to 1D All-Reduce. 
When comparing 1D collective versus 2D collective, we can see a significant reduction in completion time from \autoref{fig:DLRM_CT} for all the congestion control policies. Also, from \autoref{fig:DLRM_pfc}, we also observe a significant reduction in the count of PFCs in 2D collective versus 1D collectives as we are sending less data into the network. 






\subsection{Real Workload: DLRM Results}
In this section, we show the effect of different CC algorithms on the end-to-end training time of the DLRM model described in \autoref{subsec:targetWorkload}.
\autoref{fig:workload} shows one training iteration time, decomposed into total compute and total exposed communication, for different CC algorithms and two variations of the All-Reduce algorithm (i.e., 1D and 2D). In total, each training time of DLRM issues 109.5 MB All-Reduce and 8 MB All-To-All for MLP and embedding layers, respectively. 

As can be seen, choosing the hierarchical 2D All-Reduce algorithm results in much better performance due to better utilizing local High-BW NVlinks and sending less data through NICs. However, current CC algorithms show a slight variance in end-to-end performance with each algorithm. More specifically, the end-to-end performance difference is less than 4\% across all CCs, and No PFC gives the best performance results along with DCQCN, DCTCP, and HPCC-PINT. HPCC works worst in 1D and 2D All-Reduce cases mainly due to its added header overheads that reduce the \emph{network goodput}. The performance of TIMELY is not as poor as in earlier experiments, as the same parameters are almost optimal in this scenario. 

\insertFigureNew{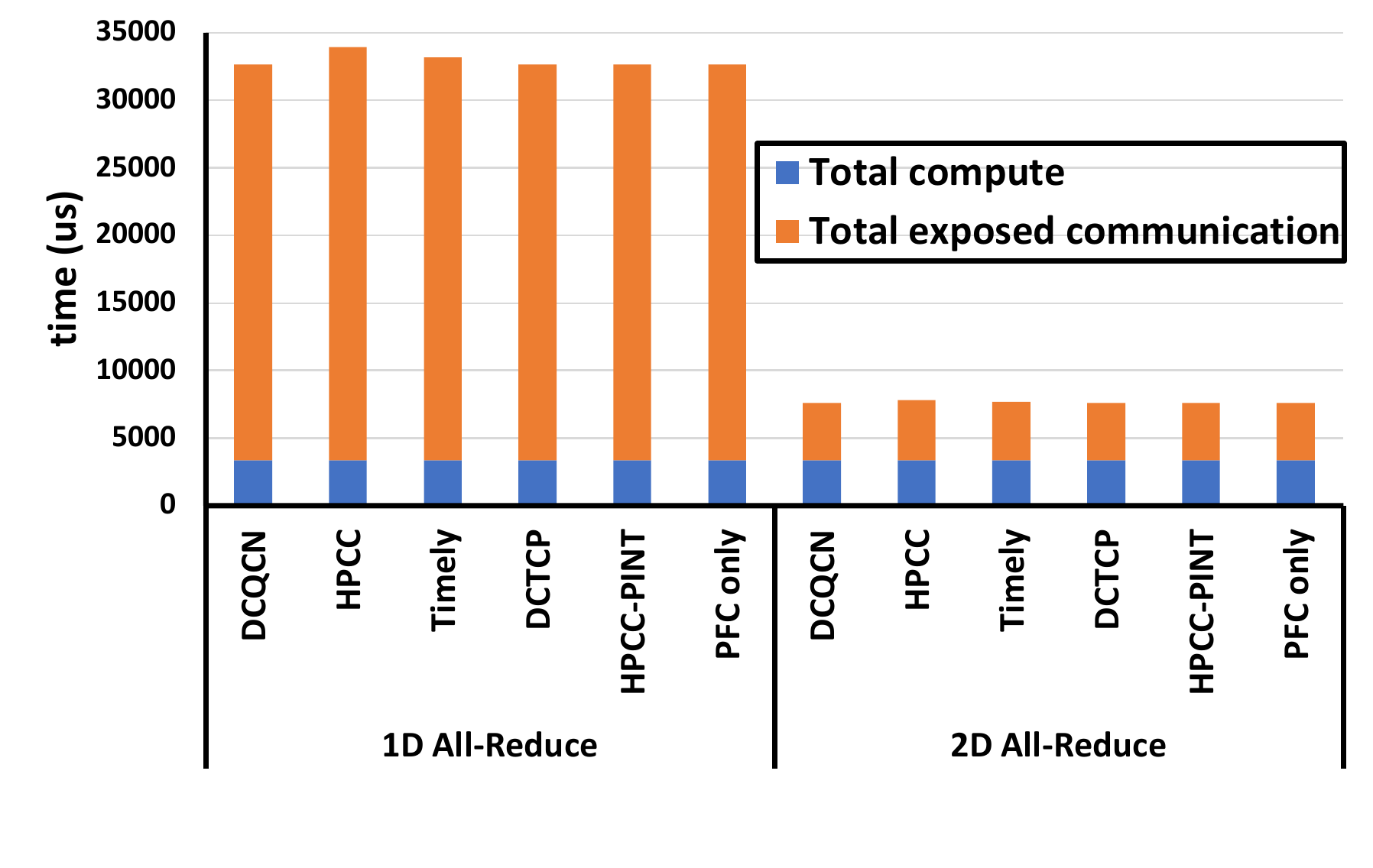}{The end-to-end total compute and total exposed communication for 1 training iteration of DLRM model running on 128$\times$V100  GPUs.}{0.8}

\subsection{Discussion}
Scalable topology-aware collective algorithms are inherently designed to avoid in-cast patterns and balance traffic optimally. Our results show that baseline PFC is equal to other proposed congestion control (CCs) in terms of end-to-end performance. This means that state-of-the-art CC algorithms introduce considerable overhead at the endpoints, requiring frequent network feedback but offering little performance improvements (for the single workload scenario). Note that compared to the general datacenters, the overhead of CC algorithms are much higher in distributed training since they use high-performance network components (e.g., 400 Gbps NICs \cite{NIC400G}) and also host multiple NICs per server node, wasting a lot of cycles at the endpoints to run CC algorithm. The only benefit of proposed CCs over the baseline PFC is that reducing the number of PAUSE frames minimizes the chance of PFC deadlocks that can rarely happen and halt the network. Additionally congestion control algorithms like DCQCN and TIMELY require extensive hyperparameter tuning and may not be optimal for all the flows. Additionally, tuning the congestion control hyperparameter before running every deep learning workload is not a feasible solution. Hence, we conclude that an optimized CC mechanism for distributed training can be much simpler than other complicated (with high overhead) CCs. The only feature optimized CC needs to provide is to prevent rare event deadlocks. This can be done by setting the congestion window to minimize buffer usage and hence minimize the PFC and chance of deadlocks. Fortunately, the communication patterns of distributed training are deterministic and repeated for each training iteration. Therefore an optimized CC can be a very low overhead by leveraging this deterministic communication behavior and statically setting the congestion window to minimize the chance of deadlock while obtaining the same performance as baseline PFC. Designing such CC and studying the multi-workload cases are the subject of our future studies.
\section{Related Works}
Although congestion control schemes are vastly studied for general datacenter networks, there is little research on CC algorithms for distributed training. Rashidi et al.  \cite{hotiPaper} studied different TCP network configurations for distributed training of recommendation models. Moreover, recent works have shown the applicability of reinforcement learning for internet network congestion control \cite{RLCC1}\cite{RLCC2}. In contrast, this paper mainly focuses on RoCE CC algorithms for distributed training algorithms.
\section{Conclusion}
In this paper, we provided a detailed analysis on comparing different proposed RoCE congestion control schemes for distributed training workload/platforms. We highlighted the fact that distributed training workloads/platforms have distinct characteristics that set them apart compared to the general datacenter environments. Therefore, it is essential to tune the networking stack for such platforms separately. We compared the baseline PFC against DCQCN, DCTCP, Timely, and HPCC congestion control schemes. We presented their behavior for both microbenchmark workloads and end-to-end distributed training workloads. In future, we also plan to study the effect of congestion control when multiple training workloads share the same datacenter resources. Based on our analysis, we defined the characteristics of a tuned congestion control scheme for distributed training, which is the subject of our future work. 






%

\bibliographystyle{IEEEtran}
\bibliography{references}

\end{document}